\documentclass[12pt]{iopart}
\usepackage{graphicx}
\usepackage{dcolumn}
\usepackage{bm}
\usepackage{color}
\usepackage{amssymb}
\newcommand{\beq}{\begin{equation}}
\newcommand{\ket}[1]{\left|#1\right\rangle} 
\newcommand{\bra}[1]{\left\langle#1\right|} 
\newcommand{\sand}[3]{\left\langle\,#1\left|\,#2\,\right|#3\,\right\rangle} 
\newcommand{\eeq}{\end{equation}}
\newcommand{\bpm}{\begin{pmatrix}}
\newcommand{\epm}{\end{pmatrix}}
\newcommand{\bea}{\begin{eqnarray}}
\newcommand{\eea}{\end{eqnarray}}

\newcommand{\ignore}[1]{}

\begin{document}


\title[FQH states production in gases of ultracold bosonic atoms subjected to
  geometric...]
{Fractional quantum Hall states of few bosonic atoms in geometric gauge fields}

\author{B. Juli\'a-D\'{\i}az$^{1,2}$, T. Gra\ss$^2$, 
N. Barber\'an$^1$, and M. Lewenstein$^{2,3}$}

\address{$^1$ Departament d'Estructura 
i Constituents de la Mat\`{e}ria, Facultat de F\'\i sica, \\ 
Universitat de Barcelona, E--08028, 
Barcelona, Spain}

\address{$^2$ ICFO-Institut de Ci\`encies Fot\`oniques, 
Parc Mediterrani de la Tecnologia, 08860 Barcelona, Spain}

\address{$^3$ICREA-Instituci\'o Catalana de Recerca i Estudis Avan\c cats, 
08010 Barcelona, Spain}

\vskip11mm

\medskip

\begin{abstract}

We employ the exact diagonalization method to analyze the possibility 
of generating strongly correlated states in 
two-dimensional clouds of ultracold bosonic atoms which are subjected 
to a geometric gauge field created by coupling two internal atomic 
states to a laser beam. Tuning the gauge field strength, the system 
undergoes stepwise transitions between different ground states, which 
we describe by analytical trial wave functions, amongst them 
the Pfaffian, the Laughlin, and a Laughlin quasiparticle many-body state. 
The adiabatic following of the center of mass movement by 
the lowest energy dressed internal state, is lost by the mixing of the 
second internal state. This mixture can be controlled by the intensity of 
the laser field. The non-adiabaticity is inherent to the considered 
setup, and is shown to play the role of circular asymmetry. We study 
its influence on the properties of the ground state of the system. 
Its main effect is to reduce the overlap of the numerical 
solutions with the analytical trial expressions by 
occupying states with higher angular momentum.
Thus, we propose generalized wave 
functions arising from the Laughlin and Pfaffian wave function by including 
components, where extra Jastrow factors appear, while preserving 
important features of these states. We analyze quasihole excitations 
over the Laughlin and generalized Laughlin states, and show that they 
possess effective fractional charge and obey anyonic statistics. 
Finally, we study the energy gap over the Laughlin state as the number of particles is 
increased keeping the chemical potential fixed. The gap is found to 
decrease as the number of particles is increased, indicating that the 
observability of the Laughlin state is restricted to a small number of particles.

\end{abstract}
\vskip2mm

\maketitle

\section{Introduction}
\label{sec1}
A spectacular progress on the manipulation and control of cold 
atomic clouds has been achieved since the first experimental 
realization of a Bose-Einstein condensed system by Anderson 
{\it et\, al.}~\cite{and} and Davis {\it et\, al.}~\cite{dav} 
nearly simultaneously in $1995$. On the one hand, these systems 
provide us with a toolbox for studying the principles of quantum 
mechanics, as has for instance been done in experiments involving 
the interference of two condensates leading to the manifestation 
of coherence~\cite{and2}. On the other hand, cold atoms have been 
used in order to simulate interesting phenomena appearing in 
condensed-matter physics~\cite{Lewenstein07}, like the 
Mott-insulator to superfluid transition~\cite{gre}.

A severe restriction, however, stems from the atom's 
electro-neutrality, which hinders a direct implementation 
of phenomena involving electromagnetic forces. An important 
example of the latter is the physics of the integer and 
fractional quantum Hall effect, occurring in two-dimensional 
interacting systems 
under the presence of a strong perpendicular magnetic 
field. The ground states of fractional quantum Hall (FQHE) systems 
are highly correlated states, like the Pfaffian state 
proposed by Moore and Read~\cite{moo}, or the celebrated 
Laughlin state~\cite{lau}, whose bosonic analogs are found 
to be the exact eigenstates of a Hamiltonian with a 
three-body (3B), or two-body (2B)~\cite{wil,coo} contact 
interaction, respectively. In addition, the quasihole 
excitations over the Laughlin state have fractional 
effective charge and fractional statistics~\cite{arovas}. 
Excitations of the Pfaffian state may even obey 
non-Abelian braiding statistics~\cite{nawi,rezayi}, which 
makes them interesting from both fundamental and technological 
points of view~\cite{nayak}.

Several proposals of experimental routes to obtain these types 
of states have appeared in the literature, ranging from the 
use of rotating traps to simulate magnetic fields acting 
on charges to the use of laser-beam configurations acting on 
atoms with several internal states~\cite{mad,par2,pop,blo,ken,ron1,ron2,jul}.    
In this article we analyze the appropriate conditions to realize 
some strongly correlated states within a simple configuration of 
a single laser beam shining on a cloud of cold atoms with two 
internal states. If the internal dynamics of the atoms, governed 
by the Rabi frequency of the atom-laser coupling, is fast enough 
with respect to the slow variation of center of mass position, 
then the atoms evolve adiabatically. They remain in one definite 
space-dependent superposition of the internal bare states, and 
the accumulation of Berry phase~\cite{zee84} during its movement mimics an 
effective magnetic field~\cite{litu1,litu2}. An important goal is to go further and 
analyze the effect of a slight amount of non-adiabaticity, which 
we treat in a perturbative way. 

Through the controlled variation of external parameters, different 
strongly correlated states appear in the spectrum, i.e. a
Laughlin-like state, a Pfaffian-like state and the quasiparticle-like 
state obtained from the Laughlin. Our main aim is to map the 
regions in parameter space where the exact, numerically obtained, 
ground state (GS) has a large overlap with explicit analytical expressions 
provided for these relevant strongly correlated states. To remain 
close to possible experimental implementations, we study the 
effect of small perturbations that create non-adiabaticities which 
are unavoidable for finite values of the laser intensity. The 
small perturbation produces a deformation of the atomic cloud 
preserving most of the notable 
properties of the original states, e.g. entropy, internal energy, or  
anyonic character of excitations. These slightly asymmetric ground 
states are well represented by generalized analytic wave functions.    

Focusing additional laser beams on the atomic cloud, it is possible to  
pierce holes into the system. As long as the asymmetric 
perturbation is small, the resulting states can be well 
described by an analytical quasihole wave function, which can 
be obtained from the Laughlin or the generalized Laughlin wave 
function. In both cases, the effective charge and statistical 
phase of the quasihole excitations are found to attain fractional 
values, demonstrating the possibility of observing anyons within 
the proposed setup.

In this work we concentrate on the physics of few-body systems, 
independently of their attainability in the thermodynamic 
limit, which is beyond the scope of the present article. 
This approach is meaningful as there are nowadays a number 
of groups able of dealing with small bosonic clouds using 
several techniques~\cite{kino,gemelke}. In particular, 
Ref.~\cite{gemelke} has presented experimental evidence for 
the production of quantum states of fractional quantum Hall 
type for small atom systems (N$<$10). These experimental 
developments have triggered a number of theoretical proposals 
focusing on the production of strongly correlated quantum states 
in small atomic clouds~\cite{pop,ron1,ron2}. We analyze the 
dependence on the number of particles of some of our main results. 
First, we show that observing the anyonic character of quasihole 
excitations becomes possible when $N\gtrsim 5$. Second, we show 
that at fixed chemical potential the bulk gap of the Laughlin 
state is decreased as $N$ is increased, indicating that 
few-body systems provide the best scenario for producing the 
bosonic Laughlin state.

The paper is organized as follows. In Section~\ref{sec2} we present 
the system and derive an effective Hamiltonian to describe it. In 
Section~\ref{sec3} we provide the analytical expressions of the 
relevant fractional quantum Hall states and their generalizations. 
Our results are presented in Section~\ref{sec4} and Section~\ref{sec5} 
in the adiabatic/symmetric and non-adiabatic/asymmetric cases, respectively. In 
Section~\ref{sec8} we study the fractional charge and anyonic 
statistics of quasiholes, produced by means of additional lasers. 
In Section~\ref{sec6} we analyze the behavior of the energy gap 
above the Laughlin state as $N$ increases, and its evolution as 
a function of the system's parameters. Finally, in Section~\ref{sec7}, 
we present our conclusions. 


\section{Description of the system}
\label{sec2}


We consider a setup to produce artificial gauge fields in a 
small cloud of ultracold atoms closely following the configuration 
described in Ref.~\cite{jul}. The system is confined by harmonic 
traps. The confinement in the $z$ direction is assumed to be 
strong enough to achieve effectively a two dimensional system. 
The cloud is illuminated by a single laser beam with wave number 
$k$ and frequency $\omega_L$, which propagates in the $y$-direction 
and is close to the resonance with a transition between two internal atomic 
states, $\omega_L=\omega_A$. The interaction between the electric 
field of the laser and the induced electric dipole is modeled 
by the atom-laser Hamiltonian, which in the rotating-wave 
approximation and in the rotating frame is given by~\cite{coh,bas},
\begin{equation}
\label{HAL}
H_{\rm AL}= 
E_{\rm g}\ket{\rm g}\bra{\rm g} 
+ (E_{\rm e}-\hbar \omega_L) \ket{\rm e}\bra{\rm e}
+\frac{\hbar\Omega_0}{2}{\rm e}^{iky}\ket{\rm e}\bra{\rm g}
+\frac{\hbar\Omega_0}{2}{\rm e}^{-iky}\ket{\rm g}\bra{\rm e}       
\end{equation}
\medskip
where $E_{\rm g}$ and $E_{\rm e}$ are the energies of the bare 
atomic ground and excited state and $\Omega_0$ is the Rabi 
frequency, which is proportional to the laser intensity. No 
spontaneous emission of photons from the excited state is 
considered. This assumption is justified as long as the 
lifetime of the excited state is longer than the typical 
duration of an experiment. Lifetimes of several seconds, 
as found for Ytterbium or some Alkaline earth metals, should be 
sufficient.

In order to obtain a non-trivial gauge potential from the 
coupling in Eq.~(\ref{HAL}), we still have to make it dependent 
on $x$. This can be achieved via a real magnetic field, which 
by Zeeman effect makes the internal energy levels vary linearly 
with $x$. Introducing a parameter $w$, setting the length 
scale of this shift, we have, 
\beq
E_{\rm g}=-\frac{\hbar\Omega_0}{2}\frac{x}{w}\,, \qquad 
E_{\rm e}=\hbar\omega_A+\frac{\hbar\Omega_0}{2}\frac{x}{w}\,\,.
\eeq
Then, the single particle Hamiltonian is given by
\beq
H_{\rm sp} =\frac{p^2}{2M}+V(\vec{r})+\frac{\hbar \Omega}{2}
\left(\matrix{
\cos\theta & {\rm e}^{i\phi} \sin\theta \cr
{\rm e}^{-i\phi} \sin\theta & -\cos\theta
}\right) \, ,
\label{eqal} 
\eeq 
where the third term is the atom-laser Hamiltonian 
represented in the $\{\ket{\rm e},\ket{\rm g}\}$ basis. 
Here, $M$ is the atomic mass, $\Omega=\Omega_0\sqrt{1+x^2/w^2}$, 
$\sin \theta=w/\sqrt{w^2+x^2}$, $\cos \theta=x/\sqrt{w^2+x^2}$, 
and $\phi=ky$. $V(\vec{r})$ is the trap potential that will 
be fixed below.

Up to this point, the Hamiltonian is given by physically 
measurable parameters. In the next step, we choose particular 
expressions for the single particle states that diagonalize 
the atom-laser Hamiltonian. This corresponds in fact to 
the selection of a specific gauge for the vector and scalar 
potentials that will drive the center of mass dynamics, as 
will be shown below. We consider the eigenfunctions of 
$H_{\rm AL}$ given by,
 \beq
\ket{\Psi_1} ={\rm e}^{-iG}
\left(\matrix{
C\,\, {\rm e}^{i\phi/2}\cr
S\,\,{\rm e}^{-i\phi/2} 
}\right)
\, ,\qquad 
\ket{\Psi_2} ={\rm e}^{iG}
\left(\matrix{
-S\,\, {\rm e}^{i\phi/2}\cr
C\,\,{\rm e}^{-i\phi/2} 
}\right)
\, ,
\label{eq4}
\eeq 
in the $\{\ket{\rm e},\ket{\rm g}\}$ basis, where $C=\cos{\theta/2}$,
$S=\sin{\theta/2}$, and  $G=\frac{kxy}{4w}$. The atomic state can be 
then expressed as,
\beq
\chi(\vec{r})=a_1(\vec{r})\otimes \ket{\Psi_1}+a_2(\vec{r})\otimes \ket{\Psi_2}
\eeq
where $a_i$ captures the external dynamics and $\ket{\Psi_i}$ provides 
internal degree of freedom. The single-particle Hamiltonian is then 
expressed in the $\{\ket{\Psi_1},\ket{\Psi_2}\}$ basis as
\beq
H_{\rm sp} =
\left(\matrix{
H_{11} & H_{12}\cr
H_{21}& H_{22}
}\right)\,,
\label{eq6}
\eeq 
acting on the spinor $[a_1(\vec{r}),a_2(\vec{r})]$. Defining a 
vector potential $\vec{A}$,
\begin{eqnarray}
\vec{A}(\vec{r}) = 
\hbar k \left[ \frac{y}{4w} , \frac{x}{4w}-\frac{x}{2\sqrt{x^2+w^2}}
\right]
\end{eqnarray} 
and a scalar potential $U$,
\begin{eqnarray}
 U(\vec{r}) = 
\frac{\hbar^2 w^2}{8M\left(x^2+w^2\right)}\left(k^2+\frac{1}{x^2+w^2}\right)\,,
\end{eqnarray}
we obtain,
\beq
H_{11}=\frac{\left[ \vec{p} - \vec{A}(\vec{r}) \right]^2}{2M}+U(\vec{r})+V(\vec{r})
+\frac{\hbar\Omega(\vec{r})}{2},
\label{eq7}
\eeq
and
\beq
H_{22}=\frac{\left[ \vec{p} + \vec{A}(\vec{r}) \right]^2}{2M}
+U(\vec{r})+V(\vec{r})-\frac{\hbar\Omega(\vec{r})}{2},
\label{eq8}
\eeq
which are the Hamiltonians driving the external dynamics of atoms being in the
internal state $\ket{\Psi_1}$ and $\ket{\Psi_2}$, respectively.
By expanding the $H_{ij}$ terms up to second order in $x$ and $y$, which is
justified by choosing $w$ to be larger than the extension of the cloud, we
find that the energy distance between these two manifolds is given 
by $\hbar\Omega_0$. 
For convenience, we make the Hamiltonian element for the low energy manifold,
$H_{22}$, independent of $\Omega_0$ by adding the constant term 
$\frac{\hbar \Omega_0}{2}$ to the diagonal of $\hat H_{\mathrm{sp}}$. Further we note
that with the explicit selection of Eq.~(\ref{eq4}) and for $x,y \ll w$,
$\vec{A}$ is in the 
symmetric gauge: 
$\vec{A} \approx \frac{\hbar k}{4w}(y,-x)$. This allows for making $H_{22}$
fully symmetric by a proper choice of the trapping frequency. Eq.~(\ref{eq8}) 
then takes the form
\beq
H_{22} =\frac{p^2}{2M}+\frac{\vec{p}\cdot \vec{A}}{M}+\frac{M}{2}\omega_{\perp}(x^2+y^2)\,,
\eeq 
and we can take the advantage from the knowledge of its single particle eigenfunctions, namely 
the Fock-Darwin states~\cite{jac}. This final expression is formally equal to the
Hamiltonian driving a system of charges trapped by a
harmonic potential of frequency $\omega_{\perp}$,
under a constant magnetic field along the $z$-direction, or 
equivalently, a system of neutral atoms trapped by a 
rotating harmonic potential of frequency $\omega_{\perp}$, 
expressed in the rotating frame of reference~\cite{Cooper:2008,Fetter:2009}. 

While the equivalence to the rotating case holds for $H_{22}$, we stress that
it does not for the system's behavior in the upper manifold described by
$H_{11}$. However, due to the off-diagonal terms 
in $\hat H_{\mathrm{sp}}$, both manifolds are coupled. Typical expected values
of $H_{12}$ and $H_{21}$ 
are of the order 
of the recoil energy $E_R=\frac{\hbar^2k^2}{2M}$ which gives the 
scale for the kinetic energy of the atomic center-of-mass motion 
when it absorbs or emits a single photon. If we consider $\hbar \Omega_0\gg E_R$, this 
coupling is small, and we can restrict ourselves to the low 
energy manifold. Namely, we are in the situation where the 
internal dynamics is much faster than the center of mass motion 
and can follow the external variations in a quasiadiabatic way~\cite{Dalibard:2011}. 

To go beyond the adiabatic approximation, we consider the influence of 
the high-energy manifold as a small perturbation. Using the procedure 
appropriate to systems that show two significantly different energy 
scales as explained in Ref.~\cite{coh}, and detailed in~\ref{apb},
we calculate an effective Hamiltonian up to second order in the perturbation, which reads 
\beq
H_{22}^{\rm eff} =H_{22}-\frac{H_{21}H_{12}}{\hbar\Omega_0},
\label{eq10}
\eeq 
where the explicit expression for the perturbative term $H_{21}H_{12}/\Omega_0$
is given in ~\ref{apb}.
Though mathematically more involved and physically richer, this term 
is reminiscent of the anisotropic potential 
that is 
applied to set an atomic cloud in rotation. Usually, the expression used to
model the stirring laser 
is given by  
$\alpha M \omega_{\perp}(x^2-y^2)$~\cite{par,dag}, where $\alpha$ measures the
strength of the deformation. Similarly the term $H_{21}H_{12}/\hbar\Omega_0$
can only produce changes of $L$ of $\Delta L= 0, \pm2, \pm 4$
(see~\ref{apb}). In what follows we will identify 'deformation' with
'coupling' indicating that a large connection with $H_{11}$ implies
deformation since $H_{11}$ is not 
cylindrically symmetric. 

The many-body Hamiltonian which we finally will deal with is given by adding a
contact interaction term to the effective 
Hamiltonian from Eq.~(\ref{eq10}):
\beq
H =\sum_{i=1}^N H_{22}^{\rm eff} (i) + \frac{\hbar^2g}{M}\sum_{i<j}\delta(\vec{r}_i-\vec{r}_j)
\label{eq9}
\eeq 
where $g$ is a dimensionless parameter fixing the contact 
interaction strength. From now on we will consider $gN=6$ in 
the numerical calculations. The Hamiltonian $H$ acts only on 
the low energy manifold, which effetively is perturbed by the second manifold. 
The important two parameters in $H$ are the dimensionless ratio 
$\eta \equiv \frac{\hbar k}{4Mw \omega_{\perp}}$ and the degree of 
the perturbation given by $\Omega_0$. The expression $\eta \omega_{\perp}$ 
plays the role of the rotating frequency, and also the effective 
magnetic field strength $B_0$ is proportional to $\eta$:
\begin{eqnarray}
\label{B0}
B_0 \equiv \frac{\hbar k}{2w} = 2 M \omega_{\perp} \eta.
\end{eqnarray}
The largest possible value of $\eta$ is given by 1 in order to 
keep the system confined. 
We consider that the artificial magnetic field is strong enough to work 
exclusively in the lowest Landau level regime. To this end, 
our parameters must fulfill the following conditions: the 
energy difference between Landau levels is larger than both, 
the kinetic energy of a single particle within a Landau level 
and the interaction energy per particle. Note that the strength 
of the atom-laser coupling, characterized by $\Omega_0$, is 
different from the strength of the magnetic field, characterized 
by $\eta$. Because the magnetic field has a geometric origin, 
$\eta$ is independent of the atom-laser coupling, as long 
as the adiabatic approximation is justified.

\section{Analytical many-body states}
\label{sec3}

In this section we give an overview of the analytical wave 
functions discussed in the literature which will turn out to 
be relevant for describing our system. 

\subsection{Laughlin state} 

The well-known Laughlin state has the analytical 
form~\cite{lau,Cooper:2001,Regnault:2003},
\begin{equation}
\Psi_{\cal L}(z_1,\dots,z_N) =
{\cal N_L}\prod_{i<j}(z_i-z_j)^{1/\nu}
{\mathrm e}^{-\sum \mid z_i\mid^2/ 2\lambda_{\perp}^2}\, ,
\label{laughwave}
\end{equation}
where $\cal N_L$ is a normalization constant, $z=x+iy$ and 
$\lambda_{\perp}=\sqrt{\hbar/M\omega_{\perp}}$. The inverse of 
the exponent, $\nu$, fixes both the density of the system and 
the symmetry of the wave function. For bosons, $1/\nu$ must 
be even. The Laughlin regime discussed here will be at half 
filling, so for the rest of this paper we set $\nu=\frac{1}{2}$. 

The analysis of the squared overlap $|\langle \Psi_{\cal L}| {\rm GS}\rangle|^2$ 
of the Laughlin state with the exact ground state as a function of the 
artificial magnetic field strength $\eta$ and the atom--laser 
coupling $\Omega_0$ shows that the overlap gets reduced as 
$\hbar\Omega_0/E_R$ decreases, even for large values of $\eta$. 
Larger overlap with the ground state can be obtained by adding 
an admixture of the Laughlin state with additional Jastrow factors 
that  allow for an increase of total angular momentum, which in
the Laughlin state is given by the integer $L=\frac{N(N-1)}{2\nu}$.
Based on these observations, in Ref.~\cite{jul} an analytical ansatz 
for the ground state in the Laughlin-like region~\footnote{Defined as the parameter 
domain where $\langle L \rangle \geq N(N-1)$.} was proposed,
\beq \Psi_{\cal GL} =
\alpha \Psi_{\cal L} 
+ 
\beta \Psi_{{\cal L}1} 
+ 
\gamma \Psi_{{\cal L}2} \, ,
\label{newl} 
\eeq
from here on referred to as the generalized Laughlin (GL) state, 
with $\Psi_{{\cal L}1} = {\cal N}_1\, \Psi_{\cal L} \cdot \sum_{i=1}^N z_i^2$, 
$\Psi_{{\cal L}2}= {\cal N}_2\, \big(\tilde{\Psi}_{{\cal L}2}
- \langle \Psi_{{\cal L}1}|\tilde{\Psi}_{{\cal L}2}\rangle 
\Psi_{{\cal L}1}\big)$, and 
$\tilde{\Psi}_{{\cal L}2}= \tilde{{\cal N}}_2 \,\Psi_{\cal L} \cdot 
\sum_{i<j}^N z_iz_j$, such that we ensure 
$\langle \Psi_{\cal L}|\Psi_{{\cal L}i}\rangle=0$ and 
$\langle \Psi_{{\cal L}i}|\Psi_{{\cal L}j}\rangle=\delta_{ij}$.  
This ansatz involves components of angular momentum $L=N(N-1)$ and 
$L=N(N-1)+2$, and yields zero contact interaction energy. 
The values of $\alpha$, $\beta$ and $\gamma$ are computed as 
$\alpha=\langle \Psi_{{\cal L}} | {\rm  GS}\rangle/\sqrt{\cal N}$, 
$\beta=\langle  \Psi_{{\cal L}1} | {\rm  GS}\rangle/ \sqrt{\cal N}$, and 
$\gamma=\langle  \Psi_{{\cal L}2} | {\rm  GS}\rangle/\sqrt{\cal N}$, 
with ${\cal N}=
|\langle \Psi_{{\cal L}} | {\rm  GS}\rangle|^2 +
|\langle \Psi_{{\cal L}1} | {\rm  GS}\rangle|^2 +
|\langle \Psi_{{\cal L}2} | {\rm  GS}\rangle|^2$.

\subsection{Pfaffian (Moore-Read) state}

While the Laughlin and the generalized Laughlin state turn out 
to be good trial states for strong magnetic fields $\eta \lesssim 1$, 
for smaller field strengths the Laughlin quasiparticle (LQP) state 
and the Pfaffian (Pf) state become relevant. The Pfaffian state 
has $L=N(N-2)/2$ for even $N$, and $L=(N-1)^2/2$ for odd $N$ 
and its analytical expression reads, 
\beq
\Psi_{\cal P}=  {\cal N}_{\rm pf} {\rm Pf}([z])  \prod_{i<j}(z_i-z_j)  \,
\label{pfaf}
\eeq
with ${\cal N}_{\rm pf}$ a normalization constant, and
\beq
{\rm Pf}([z])= {\cal A} \left[ {1 \over (z_1-z_2)} {1 \over (z_3-z_4)} 
\cdots {1 \over (z_{N-1}-z_N)}\right]\,,
\eeq
where ${\cal A}$ is an antisymmetrizer of the product. As explained in
Ref.~\cite{wil,coo}, the Pfaffian state can also be computed as, 
\beq
\Psi_{\cal P}= {\cal S} 
\prod_{i<j \in \sigma_1} (z_i-z_j)^2
\prod_{k<l \in \sigma_2}(z_k-z_l)^2
\eeq
where $\sigma_1$ and $\sigma_2$ are two subsets containing $N/2$ particles 
each if $N$ is even, and $(N+1)/2$ and $(N-1)/2$ if $N$ is odd. ${\cal S}$ 
symmetrizes the expression. The Pfaffian state has been shown to be the 
lowest energy eigenstate of a Hamiltonian which contains only three-body 
contact interactions~\cite{moo,ron2}, 
\beq
H_{\rm int}=\sum_{i<j<k} \delta(z_i-z_j)\delta(z_j-z_k) \,,
\eeq
and, similarly to the Laughlin with the two-body interaction, it has 
zero three-body interaction energy. Remarkably, however, as shown in 
Refs.~\cite{wil,pop} for a two-body interaction Hamiltonian 
and for some values of $\eta$ there is a sizeable overlap 
between the ground state of the system and this state.

As we did before for the Laughlin state in Eq. (\ref{newl}), we can 
define a generalized Pfaffian (GPf) state as 
\beq \Psi_{\cal GP} =
\alpha \Psi_{\cal P} 
+ 
\beta \Psi_{{\cal P}1} 
+ 
\gamma \Psi_{{\cal P}2} \, ,
\label{newp} 
\eeq
with $\Psi_{{\cal P}1} = {\cal N}_{{\cal P}1}\, \Psi_{\cal P} \cdot \sum_{i=1}^N z_i^2$, 
$\Psi_{{\cal P}2}= {\cal N}_{{\cal P}2}\, \big(\tilde{\Psi}_{{\cal P}2}
- \langle \Psi_{{\cal P}1}|\tilde{\Psi}_{{\cal P}2}\rangle 
\Psi_{{\cal P}1}\big)$, and 
$\tilde{\Psi}_{{\cal P}2}= \tilde{{\cal N}}_{{\cal P}2} \,\Psi_{\cal P} \cdot 
\sum_{i<j}^N z_iz_j$. Again, the parameters $\alpha$, $\beta$, and $\gamma$
are fixed to maximize the overlap of the numerical 
ground state with $\Psi_{\cal GP}$.

\subsection{Laughlin-quasiparticle state}

The Laughlin-quasiparticle state arises from the Laughlin state by 
increasing the density at the origin, decreasing its angular momentum,   
$L_{\rm qp}=N(N-1)-N$. The latter formula holds if the quasiparticle 
is at the origin. Otherwise, it also carries angular momentum and the 
total expected value of the angular momentum of the system is no longer 
an integer. The wave function is written as, 
\beq
\Psi_{{\cal L}\rm qp}= {\cal N}_{\rm qp}(\xi,\xi^*) \;{(\partial_{z_1}-\xi})\cdots 
{(\partial_{z_N}-\xi)} \Psi_{\cal L} \,,
\label{lqp}
\eeq
with $ {\cal N}_{\rm qp}(\xi,\xi^*)$ a normalization constant that
depends on the position $\xi$ and $\xi^*$ of the excitation. 
Also for the Laughlin-quasiparticle state we define a generalized 
version (GLQP), built up from the same 
Jastrow factors used in Eq.~(\ref{newl}), i.e.
\beq \Psi_{{\cal GL}\mathrm{qp}} =
\alpha \Psi_{{\cal L}\rm qp} 
+ 
\beta \Psi_{{\cal L}\rm qp1} 
+ 
\gamma \Psi_{{\cal L}\rm qp2} \, ,
\label{newlqp} 
\eeq
with $\Psi_{{\cal L}\rm qp1} = {\cal N}_{{\cal L}\rm qp1}\, \Psi_{{\cal L}\rm qp} \cdot \sum_{i=1}^N z_i^2$, 
$\Psi_{{\cal L}\rm qp2}= {\cal N}_{{\cal L}\rm qp2}\, \big(\tilde{\Psi}_{{\cal L}\rm qp2}
- \langle \Psi_{{\cal L}\rm qp1}|\tilde{\Psi}_{{\cal L}\rm qp2}\rangle 
\Psi_{{\cal L}\rm qp1}\big)$, and 
$\tilde{\Psi}_{{\cal }\rm qp2}= \tilde{{\cal N}}_{{\cal L}\rm qp2} \,\Psi_{{\cal L}\rm qp} \cdot 
\sum_{i<j}^N z_iz_j$.

\subsection{Laughlin-quasihole state}

Alternatively to increasing the homogeneous density of the Laughlin 
state locally, one might also decrease it by piercing a hole in 
the atomic cloud. Formally this is achieved by introducing an 
additional zero into the wave function, multiplying it with 
$\prod_i (\xi -z_i)$. The resulting quasihole state pierced at $\xi$ 
reads:
\begin{eqnarray}
\label{lqh}
 \Psi_{{\cal L} \mathrm{qh}} = 
{\cal N}_{{\cal L}\mathrm{qh}}(\xi,\xi^*) 
\prod_{i=1}^N (\xi - z_i) \Psi_{\cal L} \,,
\end{eqnarray}
where ${\cal N}_{{\cal L}\mathrm{qh}}(\xi,\xi^*)$ is a normalization 
constant, which explicitly depends on the position of the 
quasihole. To test the anyonic nature of the quasiholes, we 
need to do the same operation twice, such that the presence of 
two quasiholes is described by, 
\begin{eqnarray}
\label{l2qh}
 \Psi_{{\cal L} \mathrm{2qh}} = 
{\cal N}_{{\cal L}\mathrm{2qh}}(\xi_1,\xi_1^*,\xi_2,\xi_2^*) 
\prod_{i=1}^N (\xi_1 - z_i)(\xi_2-z_i) 
\Psi_{\cal L} \,.
\end{eqnarray}
The state with one quasihole has a fixed total 
angular momentum, which is $N$ quanta above the ground state, 
if the quasihole is at the origin, $\xi=0$. For off-centered 
quasihole positions, the average angular momentum is slightly 
reduced and non-integer. The state with two quasiholes has 
an angular momentum close to $2N$ quanta above the ground state.

We may also apply the same operation to the generalized Laughlin 
state and define, 
\begin{eqnarray}
 \label{glqh}
 \Psi_{{\cal GL} \mathrm{qh}} = {\cal N}_{{\cal GL}\mathrm{qh}}(\xi,\xi^*) 
\prod_{i=1}^N (\xi - z_i) 
\left[\alpha \Psi_{\cal L} +\beta \Psi_{{\cal L}1} 
+\gamma \Psi_{{\cal L}2} \right] 
\end{eqnarray}
for the state with one quasihole, and
\begin{eqnarray}
 \label{gl2qh}
 \Psi_{{\cal GL} \mathrm{2qh}} &=& 
{\cal N}_{{\cal GL}\mathrm{2qh}}(\xi_1,\xi_1^*,\xi_2,\xi_2^*) \nonumber \\
&\times &\prod_{i=1}^N (\xi_1 - z_i)(\xi_2-z_i)  
\left[\alpha \Psi_{\cal L} +\beta \Psi_{{\cal L}1} +\gamma \Psi_{{\cal L}2} \right] 
\end{eqnarray}
for the state with two quasiholes. As explained in Sec.~\ref{sec5}, 
we always find $\alpha, \beta \gg \gamma$, thus in practice we will 
consider always $\gamma\equiv 0$.

\section{Results for the adiabatic/symmetric case, $H^{\rm eff}_{22}=H_{22}$}
\label{sec4}

In the symmetric case, in which the perturbation 
$H_{21}H_{12}/(\hbar\Omega_0)$ is not included, and for 
$N=4$, four distinct regions are detected depending on the 
value of $\eta$ as previously obtained in Ref.~\cite{wil,pop}: 
Condensed, Pfaffian, Laughlin-quasiparticle and Laughlin regime. We analyze
them in the following Figs.~\ref{fig0}~and~\ref{gap}.  

The first region corresponds to a fully condensed system with 
zero angular momentum and vanishing one-body entanglement 
entropy~\footnote{The entropy is defined here from the 
one-body density matrix, as $S=-\sum n_i \ln (n_i)$, 
where $n_i$ are the eigenvalues of the one-body density matrix. 
For a more detailed discussion of the entropy we refer for 
instance to ~\cite{jul}.}, see Fig.~\ref{fig0}. The ground 
state can be well described by a wave function 
given by $\Psi_C={\cal N}_C e^{-\sum_i z_i^2/2\lambda_{\perp}}$ being 
${\cal N}_C$ a normalization constant.

\begin{figure}[t]
\begin{center}
\includegraphics*[width=100mm]{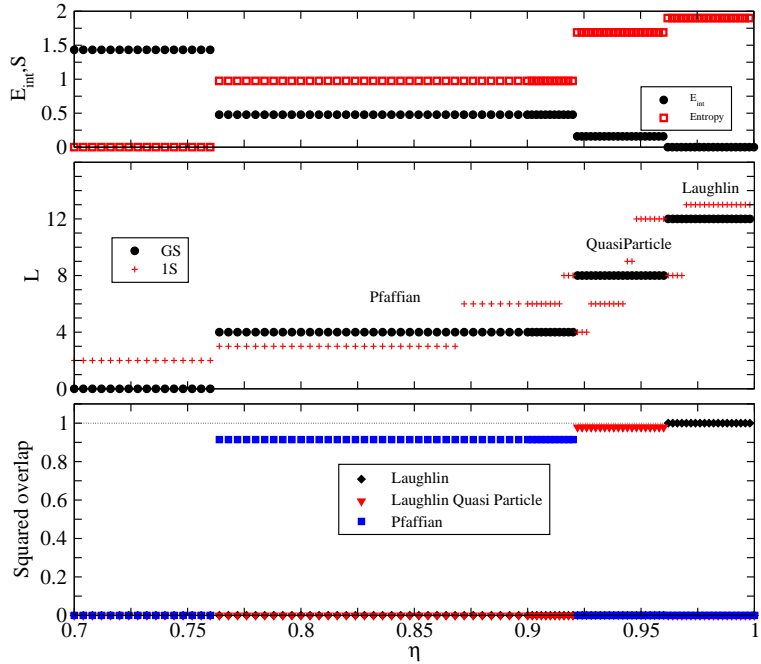}
\end{center}
\caption{(Color online) 
(upper panel) Interaction energy in units of $\hbar \omega_\perp$ 
(black circles) and one-body entanglement entropy (red squares) 
of the ground state as a function of $\eta$. 
(middle panel) Angular momentum in units of $\hbar$ of the ground state and of 
the first excited state as a function of $\eta$. 
(lower panel) Squared overlap between the ground state of 
the system and the exact Laughlin, Pfaffian and Laughlin-quasiparticle 
states. The plots corresponds to the case $H^{\rm eff}_{22}=H_{22}$.}
\label{fig0}
\end{figure}

Figs.~\ref{fig0} (middle panel) and~\ref{gap} show that the first 
excitation is of quadrupolar character up to $\eta=0.7$. Lower values 
of $\eta$ lie beyond our lowest Landau level approximation, where a 
larger Hilbert space including more Landau 
levels has to be considered.

At the critical value $\eta_1=1-gN/(8\pi) \sim 0.76$, ($gN=6$), 
a degeneracy between states with $L=0,\,2,\,3,\,$ and $4$ occurs, 
see Fig.~\ref{gap}. At this $\eta_1$ a state with broken 
symmetry, combination of several $L$-eigenstates, is precursor 
of the nucleation of the first vortex state. For increasing $\eta$
the ground state recovers the cylindrical symmetry and the angular 
momentum is $L=4$. All this phenomenology can be inferred from 
the Yrast line displayed in Fig.~\ref{yrast}. The Yrast line is 
constructed by plotting the interaction energy contribution of 
the lowest energy state for each $L$. From this line, the addition 
of the kinetic energy, which reads (up to a term independent of $L$
and $\eta$) $E_{\rm kin}= (1-\eta)L \hbar\omega_\perp$, produces the 
total energy with its minimum at the angular momentum of the ground 
state, $L_{GS}$, as exemplified for $\eta=0.85$ and $0.94$ in 
Fig.~\ref{yrast}. This is a general behavior for any $N$.

\begin{figure}[t]
\begin{center}
\includegraphics*[width=120mm]{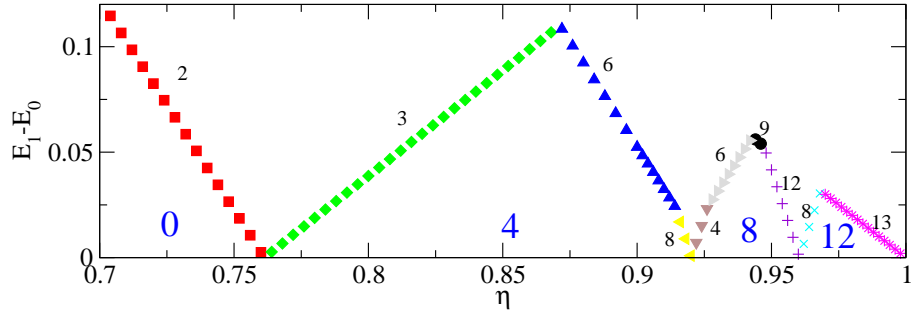}
\end{center}
\caption{(Color online) 
Energy difference, in units of $\hbar \omega_\perp$, between the 
ground state and its first excitation as a function of $\eta$. 
The large blue numbers correspond to the value of $L$ for the 
ground state. The small numbers quote the value 
of $L$ of the first excited state. $H^{\rm eff}_{22}=H_{22}$. }
\label{gap}
\end{figure}

As $\eta$ is increased above $\eta_1$, the ground state with $L=4$ 
becomes correlated as marked by the sizeable entanglement entropy 
(see Fig.~\ref{fig0}), meaning that it is not fully condensed. 
Remarkably, its squared overlap with the Pfaffian state is much 
larger (about $0.9$, see Fig.~\ref{states}) as already noticed 
in Ref.~\cite{wil,pop}, than its squared overlap with the 
one-vortex-state (about $0.47$), given by 
$\Psi_{1\mathrm{vx}}={\cal N}_{1\mathrm{vx}}\prod_{i=1}^{N} z_i e^{-\sum_iz_i^2/2\lambda_{\perp}}$
(being ${\cal N}_{1\mathrm{vx}}$ a normalization constant). If the ground state 
in this region is modeled by the eigenstate of the one-body density 
matrix with the highest occupation which plays the role of the order 
parameter in a mean field approach, it shows a well defined vortex 
at the center of the density distribution. The phase of the order 
parameter changes by $2\pi$, indicating the existence of a vortex. 
However, this function is a poor representation of the ground state 
since the non-condensed fraction is significant. This region has three 
different kind of excitations, $L=3,\,6$ and $8$ as can be seen in 
Fig.~\ref{gap}. The latter has a large overlap with the 
Laughlin-quasiparticle state as can be seen in Fig.~\ref{states}.

For $0.92\leq\eta \leq 0.96$, the ground state has $L=8$, a higher 
entanglement entropy, and a smaller interaction energy. The ground 
state has a large overlap with the Laughlin-quasiparticle state, as 
shown in Figs.~\ref{fig0} and~\ref{states}. This quasiparticle region 
has four different excitations, $L=4, 6, 9$ and 12. The $L=9$ corresponds 
to a center of mass excitation of the system as dictated by Kohn's 
theorem~\cite{jac,caza}.

Finally, for $\eta\geq\eta_2\simeq 0.96$, the ground state wave 
function is the Laughlin wave function, with zero interaction 
energy. Its excitations are the Laughlin-quasiparticle $L=8$ and 
an a center of mass excitation, Kohn mode, with  $L=13$, whose 
analytical form is, 
$\Psi= {\cal N}(z_1+z_2+z_3+z_4) \Psi_{\cal L}\,.$

\begin{figure}[t]
\begin{center}
\includegraphics*[width=80mm]{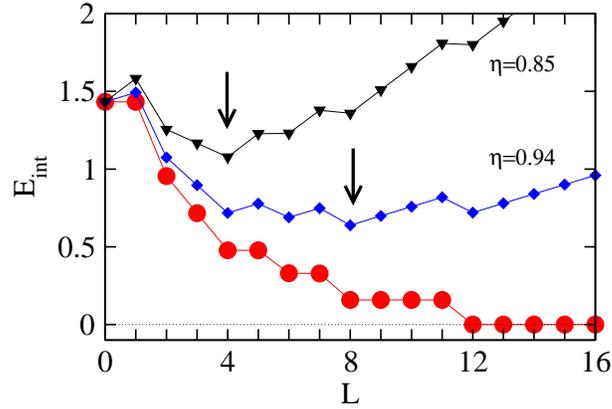}
\end{center}
\caption{(Color online) Yrast line for $N=4$, solid red circles,  
which corresponds to the interaction energy contribution of the 
lowest eigenstates for each value of $L$. The triangles and diamonds 
depict the sum of the interaction energy and the kinetic contribution 
for $\eta=0.85$ and $\eta=0.94$, respectively. The arrows mark the value 
of $L$ which corresponds to the GS in each case. The energies are given 
in units of $\hbar \omega_\perp$.}
\label{yrast}
\end{figure}

\begin{figure}[t]
\begin{center}
\includegraphics*[width=100mm]{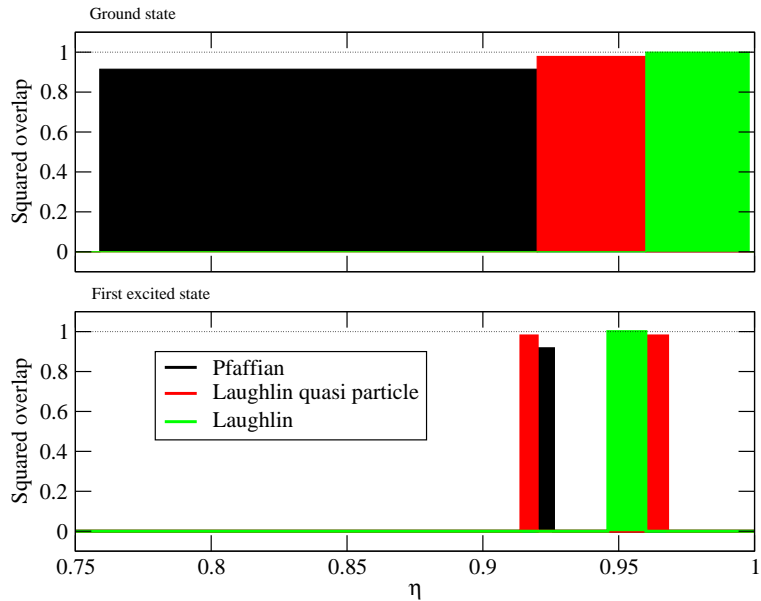}
\end{center}
\caption{(Color online) 
Squared overlap between the Pfaffian, the Laughlin-quasiparticle and 
the Laughlin states and the ground state (upper panel) and the 
first excited state (lower panel). The condensed region has been omitted. 
$H^{\rm eff}_{22}=H_{22}$. }
\label{states}
\end{figure}

\section{Effects of the non-adiabaticity/asymmetry, 
$H_{22}^{\rm eff}=H_{22}-H_{21}H_{12}/(\hbar \Omega_0)$.}
\label{sec5}

\begin{figure*}[h]
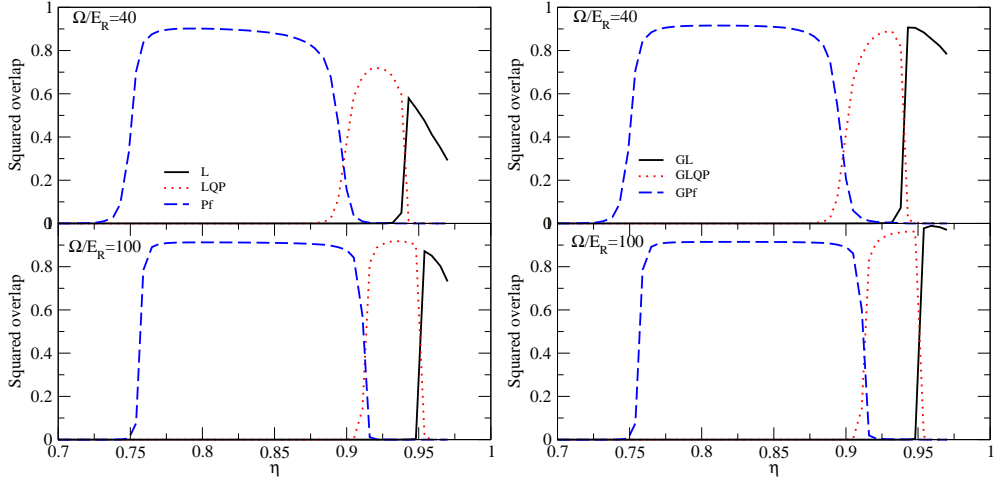

\begin{center}
\includegraphics*[width=65mm]{fig5a.eps}
\includegraphics*[width=65mm]{fig5b.eps}
\end{center}
\caption{(Color online)  (left panel) Squared overlap between 
the ground state and the original strongly correlated states 
considered, namely, the Pfaffian, Laughlin and Laughlin-quasiparticle 
states as a function of $\eta$ for $\hbar\Omega_0/E_{\rm R}=40$ and 
$100$. (right panel) Squared overlap between the GS and the 
generalized correlated states considered, GPf, GL and GLQP as 
a function of $\eta$ for $\hbar\Omega_0/E_{\rm R}=40$ and $100$. }
\label{fig1}
\end{figure*}

As discussed in Section~\ref{sec2} and~\ref{apb}, the 
considered setup can be mapped onto a symmetric Hamiltonian, 
$H_{22}$, equivalent to the one of rotating atomic clouds in 
symmetric traps, plus a term, $H_{21}H_{12}/(\hbar \Omega_0)$ 
whose importance can be controlled by tuning the laser 
coupling, $\Omega_0$. As discussed in Ref.~\cite{jul}, the first 
effect of the perturbation in the Laughlin-like region is to 
increase the angular momentum of the ground state by 
populating the states $\Psi_{{\cal L}1}$, defined in 
Eq.~(\ref{newl}). One can consider fairly small coupling 
$\hbar\Omega_0/E_R\sim 40$ and still get Laughlin-like ground 
states of the form of Eq.~(\ref{newl}), which retain most of the 
known properties of Laughlin states, namely, a large entanglement 
entropy and vanishing interaction energy~\cite{jul}. Now we extend the previous 
study to the effect of the perturbation on the Pfaffian and 
Laughlin-quasiparticle regions. 

In Fig.~\ref{fig1} we show the squared overlap between the 
ground state and the three original correlated states (left panel) 
and their generalized versions (right panel) identified as the 
generalized Pfaffian, generalized Laughlin-quasiparticle 
(GLQP) and generalized Laughlin (GL), see Eqs. (\ref{newl}), (\ref{newp}), and (\ref{newlqp}).
It turns out that in all three cases the state which is proportional to
$\gamma$ is much less populated than the states proportional to $\alpha$ and
$\beta$. We thus neglect the contribution of this term, 
for simplicity. As shown in Fig.~\ref{fig1}, overviewing all three 
regimes, the largest improvement by using the generalized versions 
occurs in the Laughlin region. Here, the total angular momentum 
increases continously with $\eta$ \cite{jul}, leading to substantial 
occupation of the state proportional to $\beta$.

\begin{figure*}[t]
\begin{center}
\includegraphics*[width=90mm]{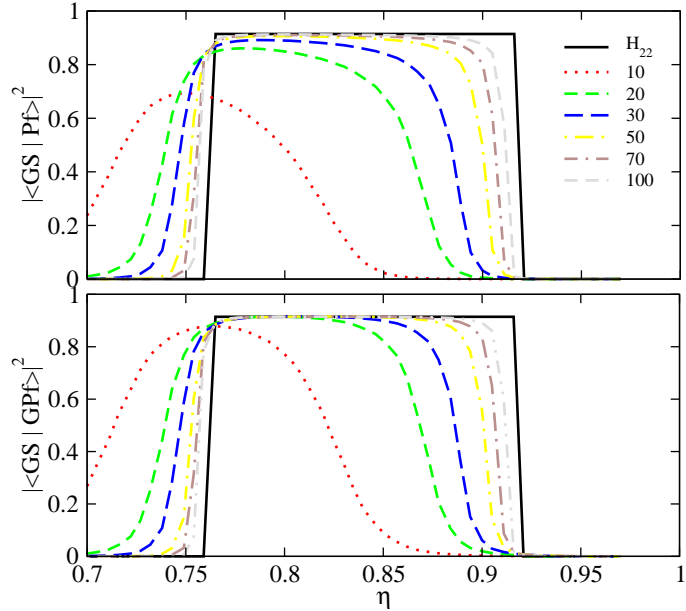}
\end{center}
\caption{(Color online)  Squared overlap between the ground state and the 
Pfaffian and generalized Pfaffian states defined in the 
text, upper and lower panels, respectively, as a function of 
$\eta$. The different lines correspond to different values of 
$\hbar\Omega_0/E_{\rm R}$. The solid line is obtained with 
$H_{22}^{\rm eff}=H_{22}$. }
\label{pfaf1}
\end{figure*}

\begin{figure*}[t]
\begin{center}
\includegraphics*[width=90mm]{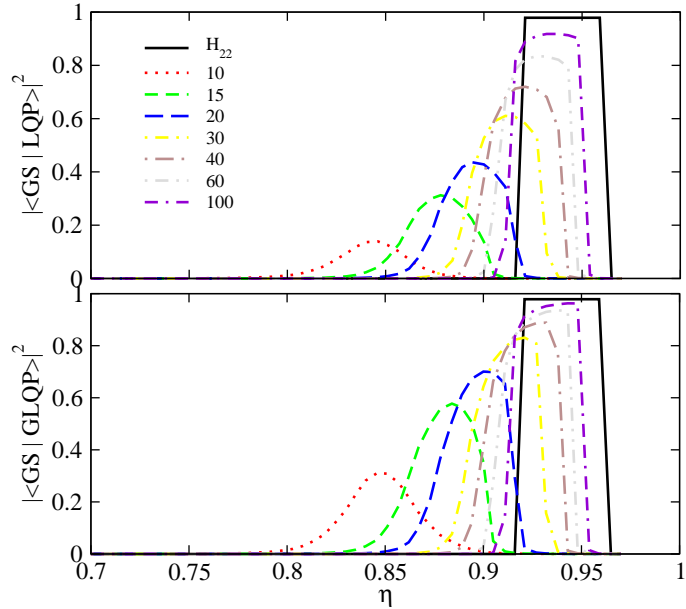}
\end{center}
\caption{(Color online)  Squared overlap between the ground state and the 
Laughlin-quasiparticle (LQP) and 
generalized Laughlin-quasiparticle (GLQP) states defined in the text, upper 
and lower panels, respectively, as a function of $\eta$. The different 
lines correspond to different values of $\hbar \Omega_0/E_R$. The solid 
line is obtained with $H_{22}^{\rm eff}=H_{22}$.}
\label{lqp1}
\end{figure*}

\begin{figure*}[tbh]
\begin{center}
\includegraphics*[width=90mm]{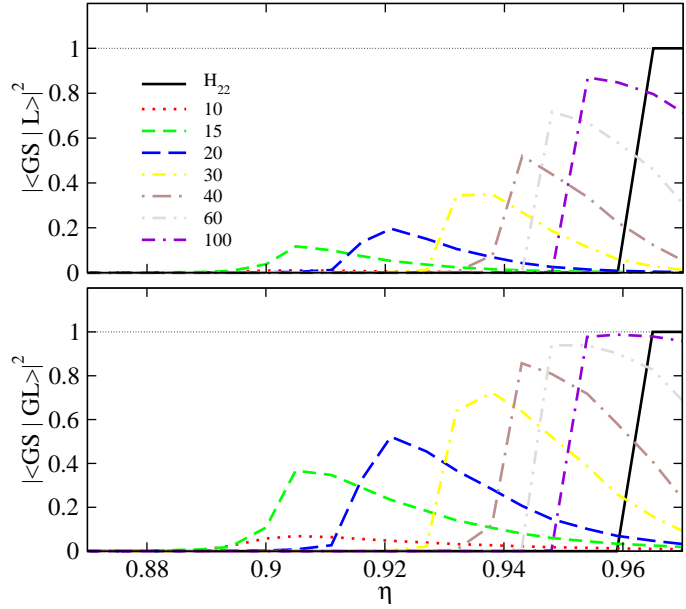}
\end{center}
\caption{(Color online)  Squared overlap between the ground state 
and the Laughlin (L) and generalized Laughlin (GL) states defined 
in the text, upper and lower panels, respectively, as a function 
of $\eta$. The different lines correspond to different values of 
$\hbar \Omega_0/E_R$. The solid line is obtained with $H_{22}^{\rm eff}=H_{22}$.}
\label{laug1}
\end{figure*}

Detailed information about the effect of the perturbation 
is shown in Figs.~\ref{pfaf1},\ref{lqp1}, and \ref{laug1} 
for each of the three regions seperately.  
First, in Fig.~\ref{pfaf1} we consider the overlap with the Pfaffian 
and GPf, exploring fairly low values of $\hbar\Omega_0/E_R$. Lower 
values of $\hbar\Omega_0/E_R$ require the consideration of higher 
order terms in the expansion of $H^{\rm eff}_{22}$ not included in our calculations.
There one can see how by decreasing the value of $\hbar\Omega_0$, 
the $\eta$ which provides maximum overlap becomes smaller. Thus, 
while in the symmetric case, the only region with non-negligible 
squared overlap with the Pfaffian was 
$0.75 \lesssim \eta \lesssim 0.92$, with, e.g $\hbar\Omega_0/E_R=20$, 
the region is roughly displaced to $0.73 \lesssim \eta \lesssim 0.89$

Also, the squared overlap with the Pfaffian gets reduced, going from 
around 0.9 in the symmetric case, to 0.7 for $\hbar\Omega_0/E_R=10$. 
As occurred with the Laughlin, the main effect of the perturbation 
is to populate states which are of the GPf type, that is, a Pfaffian 
core with appropriate Jastrow factors. The GPf state has a large 
overlap with the ground state, of the same order as the Pfaffian itself 
with the symmetric ground state. Interestingly, large values, 
$>0.8$, of the squared overlap with the GPf state can be found 
already for $\hbar\Omega_0/E_R>20$, which is relevant 
from the experimental point of view as it increases the window 
of observability.

A similar behavior is found when studying the squared overlap of 
the Laughlin-quasiparticle state with the exact ground state of 
the system. As shown in Fig.~\ref{lqp1} the region with sizeable 
overlap with the Laughlin-quasiparticle state gets shifted towards 
lower values of $\eta$ as we decrease $\Omega_0$, peaking at 
$\eta=0.85$ for $\hbar\Omega_0/E_R=10$. Also, a sizeable overlap 
is found with the GLQP state. It is however clear, that large values 
of the squared overlap, $>$0.8, can only be found for values of 
$\hbar\Omega_0/E_R>30$.

In Fig.~\ref{laug1} we present the corresponding figure for the 
case of the Laughlin and generalized Laughlin states. First, we 
note that again the region where the L and GL are most populated 
gets shifted towards lower values of $\eta$ as we decrease the 
value of $\hbar\Omega_0/E_R$. For instance, the maximum value 
is obtained around $\eta=0.91$ for $\hbar\Omega_0/E_R=15$, and this 
maximum is of 0.4 in the case of GL. Squared overlaps larger than 
0.8 are only obtained for $\hbar\Omega_0/E_R>40$. Squared overlaps 
larger than 0.5 can however be obtained with $\hbar\Omega_0/E_R$ 
as low as 20 for $\eta \sim 0.92$. 

\clearpage

\section{Fractional charge and anyonic statistics of quasihole excitations 
in the Laughlin regime}
\label{sec8}

An important property of the strongly correlated $N$-body states 
discussed in the previous sections are their excitations which might 
behave as particles with fractional charge and obey anyonic statistics, 
as is the case for the quasihole excitations over the Laughlin 
ground state~\cite{arovas,par2}. 

An experimentally feasible way of creating quasihole excitations in 
our system is by focusing a laser beam onto the atomic cloud. This 
can be described by adding the following potential to the 
single-particle Hamiltonian of Eq.~(\ref{eq9})~\cite{par2}:
\begin{eqnarray}
\label{laser}
 \hat{V}(\xi) = I \sum_{i} \delta(\xi-z_i),
\end{eqnarray}
where $I$ is the laser intensity, and $\xi$ is the position onto 
which the beam is focused. With two such potentials we should 
be able to create states with two quasiholes, according to 
Eq.~(\ref{l2qh}). In the following, we will first study quasiholes 
in the Laughlin state, which can be created when the system is in the adiabatic regime. 
Then we will also consider a slightly non-adiabatic situation, where we find
quasiholes in the GL state, as defined by Eq.~(\ref{glqh}). We analyze the
quasiholes in both the Laughlin and the GL state with respect to their 
fractional character.

\begin{figure*}[b]
\begin{center}
\includegraphics*[width=90mm]{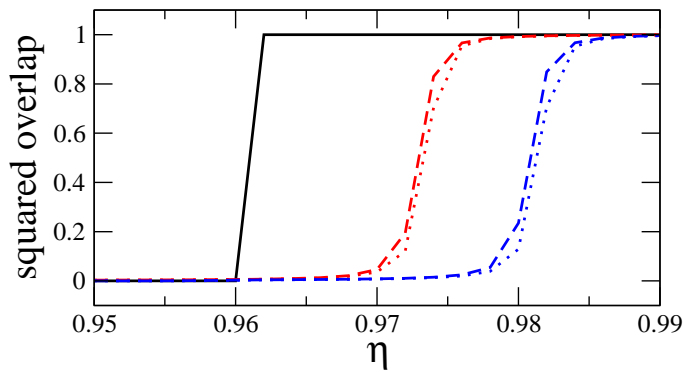}
\end{center}
\caption{(Color online) Squared overlap of exact ground states ($N=4$) with
  Laughlin state (solid), Laughlin state with one quasihole (red), and
  Laughlin state with two quasiholes (blue). The quasiholes are created by a
  laser with intensity $I=10 \frac{\hbar \omega_{\perp}}{\lambda_{\perp}^2}$
  (dotted lines) and $I=30 \frac{\hbar \omega_{\perp}}{\lambda_{\perp}^2}$ 
(dashed lines).}
\label{ov-inf}
\end{figure*}

\subsection{Quasiholes' properties in the adiabatic case}

As already discussed in Section~\ref{sec4}, for $\hbar \Omega_0 \gg E_R$, 
the system's ground state squared overlap with the Laughlin state is 
effectively one, above the critical field strength $\eta_2$. Now we 
consider the system with the additional term~(\ref{laser}) and find 
that there is also a region of $\eta$ where the overlap of the 
ground state of the system and the analytical quasihole state is 
effectively 1, see Fig.~\ref{ov-inf}. This shows that the potential 
of Eq. (\ref{laser}) is able to produce quasiholes
described by ~(\ref{lqh}). Similarly, adding two such lasers we 
also find a region of $\eta$ where the overlap between the ground 
state and the analytical state with two holes, Eq.~(\ref{l2qh}), 
is very close to 1. However, we notice that the values of $\eta$ at 
which the overlap for one or two quasiholes reaches 1 differ from 
each other, and both are found for values larger than $\eta_2$, 
see Fig. \ref{ov-inf}. These features are essentially independent of the 
laser strength $I$, for sufficiently large $I$.

\begin{figure*}[t]
\begin{center}
\includegraphics*[width=140mm]{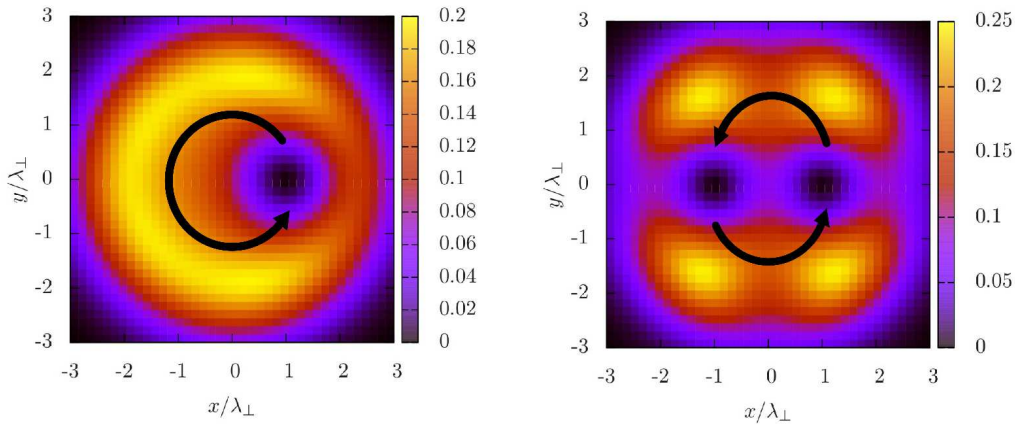}
\end{center}
\caption{(Color online) Moving one or two quasiholes at fixed radial positions
  on circles around the origin allows for determining the effective  charge
  and the statistical phase angle. The quasiholes are always fixed at a 
radial position $R=1$ (in units of $\lambda_{\perp}$).}
\label{movement}
\end{figure*}

The most interesting property of these excitations is their 
fractionality, i.e. fractional charge and statistics. To study 
the fractional charge of the quasiholes, we first note that 
in our electro-neutral system subjected to an artificial 
magnetic field, there exists the analogue of an electric 
charge which can be defined via the behavior of a particle 
or quasiparticle evolving within the gauge field. To this end, 
we consider the phase a quasihole picks up while being adiabatically 
displaced following a closed trajectory. The general expression 
for the Berry phase on a closed loop $C$ is given by~\cite{berry}
\begin{eqnarray}
\label{berryphase}
 \gamma_{\mathrm{C}} = 
i \oint_{\mathrm{C}} \bra{\Psi(\vec{R})}\nabla_{\vec{R}}\ket{\Psi(\vec{R})} 
\cdot \mathrm{d} \vec{R},
\end{eqnarray}
with $|\Psi(\vec{R})\rangle$ the state of the system, characterized 
by a parameter $\vec{R}$, which in our case is the position of the 
quasihole. For simplicity, we now assume that the quasihole is fixed 
at a radial position $|\xi|=R \lambda_{\perp}$, but is moved along a circle centered 
at the origin, see Fig.~\ref{movement}, parameterized by the angle 
$\phi$. This is sufficient to test the fractional behavior.

For general contours, one can extract the acquired phases from the
normalization factor of the quasihole state, as described in~\cite{wen}. 
For our circularly symmetric contour, however, the situation is 
simpler, as we can re-write Eq.~(\ref{berryphase}) as
\begin{eqnarray}
\gamma_{\mathrm{C}} = 
i \int_0^{2\pi} 
{ \langle \Psi_{{\cal L}\mathrm{qh}}(\phi) | \partial_\phi | 
\Psi_{{\cal L}\mathrm{qh}}(\phi)}\rangle  \mathrm{d}\phi 
\equiv \int_0^{2\pi} f(R) \mathrm{d}\phi \,.
\end{eqnarray}
Here we note that, due to the circular symmetry 
of the Laughlin state, the integrand does not depend on the 
angular position of the quasihole. The function $f(R)$ can be 
calculated by decomposing the Laughlin quasihole state into 
Fock-Darwin basis~\cite{cpc}, which we have done analytically 
for particle numbers up to 6. For compactness, we explicitly 
give here only the result for $N=4$:
\begin{eqnarray}
\label{fR}
\gamma_{\mathrm{C}} = 2\pi f(R)= 2 \pi
\frac{4 \left(10128 R^2+5313 R^4+1659 R^6+553 R^8\right) }{85572+40512
R^2+10626 R^4+2212 R^6+553 R^8} \,.
\end{eqnarray}
If we assume that the quasihole is moved sufficiently close 
to the center, i.e. $R \lesssim 1$, we can expand this 
expression in $R$ and find, 
$\gamma_{\mathrm{C}} = 2 \pi (0.473426 R^2 
+ 0.0242202 R^4 +{\cal O}(R^6)) \approx \pi R^2$. Thus, 
the acquired phase is approximately given by the enclosed area in 
units of $\lambda_{\perp}^2$. 

To obtain the effective charge of the quasihole, we must compare 
this result with the geometric phase acquired by a particle moved 
along the same closed contour due to the gauge field. In the 
Laughlin regime, where $\eta \approx 1$, we find with 
Eq.~(\ref{B0}) $B_0 \approx 2 \frac{\hbar}{\lambda_{\perp}}$, 
thus  the acquired phase $\varphi$ is two times the enclosed area in units 
of $\lambda_{\perp}^2$, 
i.e. $\varphi = \frac{1}{\hbar} B_0 (R \lambda_{\perp})^2 \pi \approx 2 R^2  \pi$.
From this follows the effective charge of the quasihole to be
$q_{\mathrm{eff}} = \frac{\gamma_{\mathrm{C}}}{\varphi} \sim 0.47$, 
close to the expected value for the Laughlin state at half filling in the thermodynamic 
limit, $1/2$~\cite{lau,arovas}.
We have performed a similar study for $N=5$ and $N=6$, finding that 
for $N=5$, the effective charge is about 0.48, and for $N=6$ it is 
found to be 0.49, i.e. by increasing the particle number the value 
1/2 is approached.

\subsubsection{Fractional statistics\\\\}

\begin{figure*}[t]
\begin{center}
\includegraphics*[width=120mm]{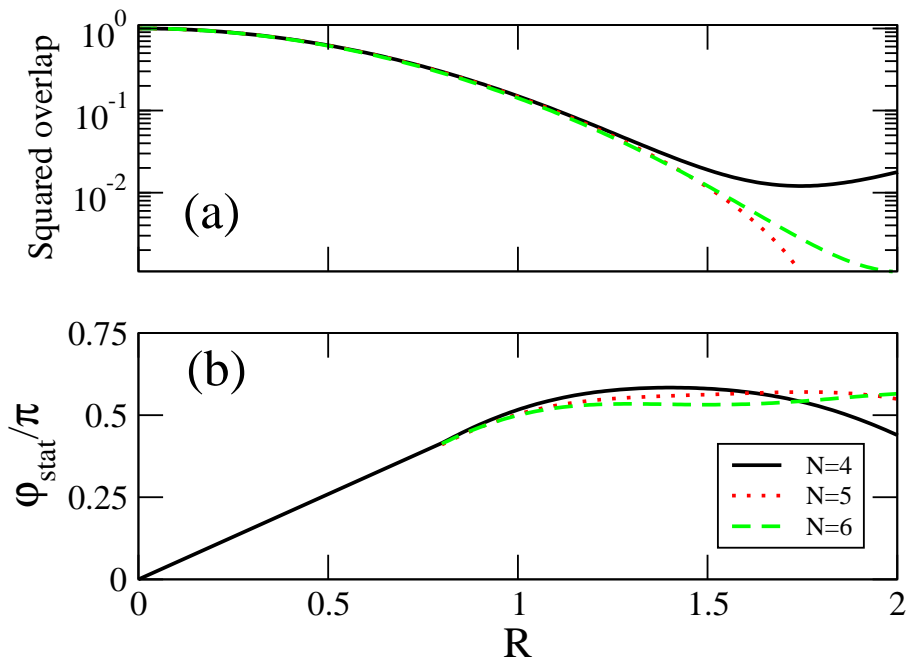}
\end{center}
\caption{
(Color online) (a) Squared overlap between the quasihole wave functions, 
$|\langle \Psi_{{\cal L} \mathrm{qh}}(\xi_1) |  \Psi_{{\cal L}
  \mathrm{qh}}(\xi_2)\rangle|^2$ at opposite angular positions, 
$\phi_1-\phi_2=\pi$, as a function their distance to the center, 
$|\xi_1|=|\xi_2|=R \lambda_{\perp}$.  
(b) Statistical angle of two quasiholes at opposite angular positions and 
radial position $R \lambda_{\perp}$. In both panels we present results for $N=4$ (black solid), 
$N=5$ (red dotted), and $N=6$ (green dashed).}
\label{statphase}
\end{figure*}

To prove the fractional statistics of the quasihole excitations we 
now consider the system with two quasiholes at $\xi_1=|\xi_1|e^{i\phi_1}$ 
and $\xi_2=|\xi_2|e^{i\phi_2}$, which we assume to sit on
opposite radial positions, i.e. $|\xi_1|=|\xi_2|=R \lambda_{\perp}$ and 
$\phi_2-\phi_1 = \pi$. We now consider the simultaneous adiabatic 
movement of the two quasiholes on two half circles, in such a way 
that, at the end, the quasiholes interchange position 
(see Fig.~\ref{movement}). This differs from a more common setup 
to test the statistical angle, where one quasihole is fixed in 
the center, while the other is encircling it, but it has the advantage 
that it maximizes the distance between the two quasiholes. Note that in
Fig.~\ref{movement}, the radial position is chosen at $R  = 1$,
i.e. the distance between the center of the quasiholes is 
$2\lambda_{\perp}$, which seems to be the minimum distance needed for 
not having a significant overlap ($<10\%$) between the two 
quasiholes, see Fig.~\ref{statphase}(a). On the other hand, in 
this small system of just four particles, larger radial positions 
lead to quasiholes overlapping with the system´s edge.

The total phase picked up during the described movement should be the 
sum of the phase picked up by one quasihole moved along a circle plus 
a phase factor due to the interchange of the two quasiholes. Again the 
phase gradient turns out to be independent from the angular position, 
but is described by a different function $\tilde f$ of the radial 
coordinate:
\begin{eqnarray}
\label{ftilde}
 \tilde f(R) = \frac{8 \left(2868120 R^4+461616 R^8+25242 R^{12}
+553 R^{16}\right)}{41660640+11472480 R^4+923232 R^8+33656 R^{12}+553 R^{16}}\,.
\end{eqnarray}
The statistical phase angle is thus, 
\begin{eqnarray}
\varphi(R)_{\mathrm{stat}} \equiv 
\int_0^{2\pi} f(R) \mathrm{d} \phi -
\int_0^{\pi} \tilde f(R) \mathrm{d} \phi 
= 2\pi f(R) - \pi \tilde f(R)\,.
\end{eqnarray}
It is shown, as a function of $R$, in Fig.~\ref{statphase} (b). First, 
as expected the statistical phase is zero if both quasiholes are in 
the same position, and it increases linearly as the distance between 
the quasiholes is increased. This linear behavior then saturates once 
the overlap between the two quasiholes, 
$|\langle \Psi_{{\cal L} \mathrm{qh}}(\xi_1) |  \Psi_{{\cal L}
  \mathrm{qh}}(\xi_2)\rangle|^2$, drops below $0.1$, and remains mostly
constant around $\pi/2$. By increasing the number of particles $N$, the 
phase angle become less dependent on $R$ once the two quasiholes do not 
overlap, meaning that it becomes a robust property of the quasiholes. 
It stabilizes around the expected value of $\pi/2$. At larger distances 
$R$, the system's edge starts to play a role.

\subsection{Non-adiabatic effects on the properties of quasiholes}

\begin{figure*}[t]
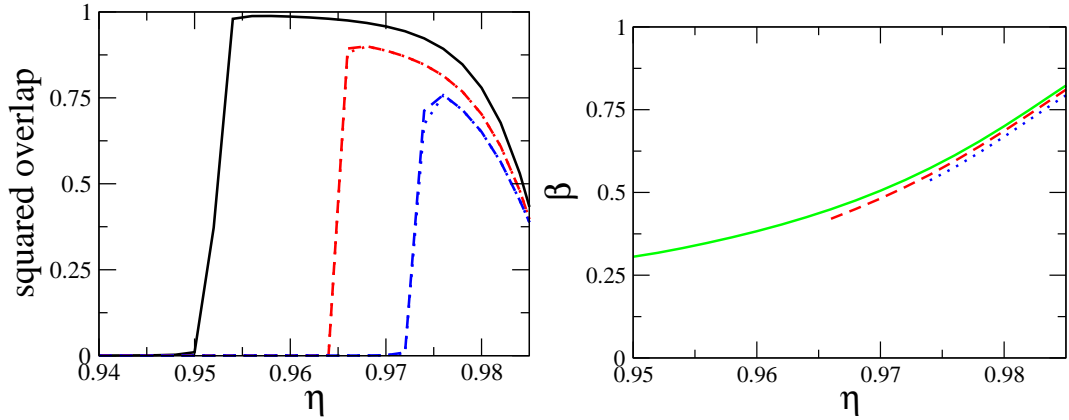

\begin{center}
\includegraphics*[width=70mm]{fig12a.eps}
\includegraphics*[width=70mm]{fig12b.eps}
\end{center}
\caption{(Color online) Left: Squared overlap of exact ground states 
($N=4$) with generalized Laughlin state (solid black), generalized 
Laughlin state with one quasihole (red), and generalized Laughlin 
state with two quasiholes (blue). The quasiholes are created by 
a laser with an intensity 
$I=30 \frac{\hbar \omega_{\perp}}{\lambda_{\perp}^2}$ (dotted lines) 
and $I=50 \frac{\hbar \omega_{\perp}}{\lambda_{\perp}^2}$ (dashed lines) 
at $\xi_1=\lambda_{\perp}$ and $\xi_2=-\lambda_{\perp}$. The Rabi 
frequency is $\hbar \Omega_0=100 E_R$. Right: The values of the 
variational parameter $\beta$ for a given $\eta$ are similar for all 
three states.}
\label{ov-100}
\end{figure*}

To study the fractionality of quasihole excitations in the non-adiabatic 
case we will again profit from the generalized analytical representations 
used to describe the ground state of the system. Following the discussion 
in the previous section we compute the squared overlap of the ground state 
obtained with no-, one-, and two-extra lasers piercing holes into the 
system. First, we find a significant squared overlap for the slightly 
perturbed case at $\hbar \Omega_0 = 100 E_R$, see Fig.~\ref{ov-100}. As 
occurred in the adiabatic case, a large overlap with the analytical one- 
and two-quasihole states appears only at higher field strengths than the 
one at which the generalized Laughlin state is reached.
Our study of the properties of quasiholes in the non-adiabatic case 
will be restricted to the parameter domain where a fair description 
of the states is provided by the generalized state given in 
Sec.~\ref{sec3}. 

We now test the behavior of the quasiholes in a generalized Laughlin
state. This can be done as before, but now we must note that the 
gradient of the state will not only depend on the radial, but also 
on the angular position of the quasiholes. Furthermore, it will 
depend on the parameter $\beta$ as defined in Eqs.~(\ref{glqh}) 
and~(\ref{gl2qh}), which is used to improve the overlap. As shown 
in Fig.~\ref{ov-100}, for given parameters $\eta$ and $\Omega_0$ 
the same value for $\beta$ optimizes simultaneously the ground 
state, the quasihole state, and the state with two quasiholes. 
We define
\begin{eqnarray}
f_{\beta}(R,\phi) \equiv 
\langle \Psi_{{\cal L}\rm qh}(\phi) |  \partial_\phi | \Psi_{{\cal L}\rm qh}(\phi)\rangle \,.
\end{eqnarray}
This function $f_{\beta}$ is quite lengthy, so we expand it in $R$ 
and give only the lowest term (${\cal O}(R^2)$):
\begin{eqnarray}
 f_{\beta}(R,\phi) \simeq
\frac {8115904 + 2799526\beta^2 -
  7102 \sqrt {94958\left (1 - \beta^2 \right)}\beta\cos(2\phi)} {17142924 +
  4477401\beta^2} R^2 
\,. &&
\end{eqnarray}
From the expression we see that for a fixed and small value of 
$R$, $f_{\beta}$ oscillates around $R^2/2$, such that the angular 
integration $\int_0^{2\pi} f_{\beta}(R,\phi) \mathrm{d}\phi$ again 
will yield a Berry phase close to the encircled area, thus half 
of the Berry phase accumulated by a normal particle.

\begin{figure}
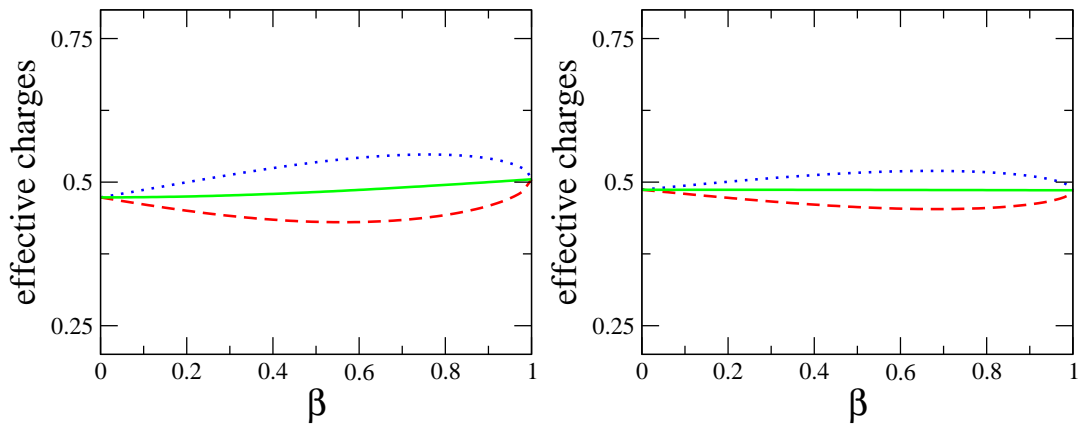

\begin{center}
 \includegraphics*[width=0.45\textwidth]{fig13a.eps}
 \includegraphics*[width=0.45\textwidth]{fig13b.eps}
\end{center}
\caption{\label{lqhc} The effective charges $q_y$ (blue dotted) and $q_x$
(red dashed), and $q_{\mathrm{eff}}=(q_x+q_y)/2$ (green solid) of quasiholes
  in the generalized Laughlin state as a function of the admixture $\beta$ of
  higher angular momentum to the 
Laughlin state, for $N=4$ (left)
 and $N=6$ (right).}
\end{figure}

Formally, we can capture this oscillating behavior by defining 
two effective charges $q_x$ and $q_y$, depending on the direction 
in which the quasihole moves. With this and by generalizing to 
arbitrary loops the Berry phase in the limit of small radial 
positions $R\lesssim 1$, the Berry phase can be written as, 
\begin{eqnarray}
 \gamma_{\mathrm{C}} 
= \oint \frac{1}{\lambda_{\perp}^2} (q_x y, - q_y x) \cdot \mathrm{d} \vec{r} 
= (q_x+q_y) \frac{A}{\lambda_{\perp}^2},
\end{eqnarray}
where $A$ is the encircled area and Stokes' theorem has been applied. 
An effective charge can be defined simply as, 
$q_{\mathrm{eff}}=(q_x+q_y)/2$. The values of $q_x, q_y,$ and
$q_{\mathrm{eff}}$ are plotted as a function of $\beta$ in Fig.~\ref{lqhc} 
for different $N$. As can be seen, in all cases the effective charges 
are close to $1/2$. For small values of $\beta$, which represent 
realistic states of the system, the value of the charges gets 
closer to $1/2$ as the number of atoms in the system is increased. 
In summary, the average charge $q_{\mathrm{eff}}$ has only a minor 
dependence on $\beta$, which decreases as $N$ increases. Though not 
realized in our system, we note that in the limit $\beta \to 1$, both charges
$q_x$ and $g_y$ again coincide due 
to the recovered cylindrical symmetry 
of the state $\Psi_{{\cal L}1}$.

\begin{figure*}[ht]
\begin{center}
\includegraphics*[width=80mm]{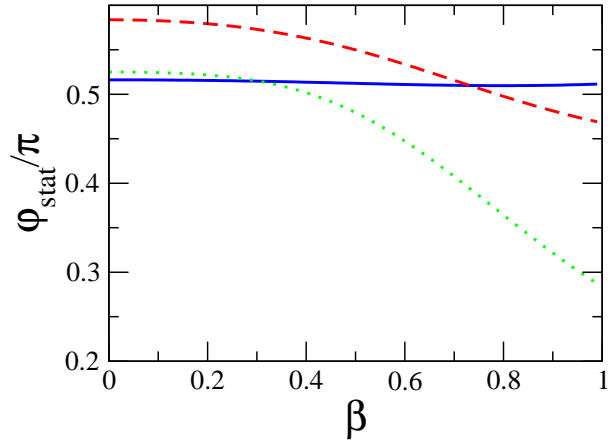}
\end{center}
\caption{(Color online) Statistical phase angle $\varphi_{\mathrm{stat}}$ for
  two quasiholes in generalized Laughlin state as a function of the
  variational parameter $\beta$. The radial position of the quasiholes is at
  $\lambda_{\perp}$ (blue solid line), $1.4 \lambda_{\perp}$ (red dashed
  line), 
and $1.8 \lambda_{\perp}$ (green dotted line).}
\label{statphase-beta}
\end{figure*}

Finally we introduce two quasiholes into the generalized Laughlin 
state. Following a procedure similar to the one for the adiabatic case presented 
in the previous section, we extract the statistical phase angle of 
the quasiholes. The result as a function of $\beta$ is shown in 
Fig.~\ref{statphase-beta} for $N=4$ and closed paths of different 
radii. While the quasiholes in the bulk, $R = 1$, remain with 
an almost constant phase angle $\varphi_{\mathrm{stat}} \approx 0.51$, 
the phase angle of quasiholes closer at the edge of the system have 
a stronger dependence on $\beta$. Thus, we find that the presence 
of a certain degree of non-adiabaticity, $\beta\leq 0.7$ does not 
spoil the presence of anyonic quasihole excitations of the Laughlin 
state.

\section{Evolution with $N$ of the energy gap}
\label{sec6}

     
The typical scenario of an experiment that has as a goal 
the realization of a specific strongly correlated state, 
is to first prepare an easily attainable initial state. Then, 
one may follow adiabatically a route in parameter space 
which ends up in the final desired state~\cite{pop,ron2}. 
This final state is expected to be robust, with a mean-life 
time larger than the time necessary to perform the 
measurements. A crucial ingredient necessary for the 
success of such approach is to have energy gaps as 
large as possible over all the ground states involved 
along the route. 

In this section we concentrate on the study of the energy 
gap over the Laughlin state. We consider first the adiabatic 
case and then study the effect of the perturbation on the 
energy gap. The gap is the energy difference between 
the ground state and the lowest excitation in the 
thermodynamic limit. We take the thermodynamic limit 
by increasing $N$ keeping the chemical potential constant during 
the process. In the homogeneous $2D$ case this corresponds 
to fixing $gN=\mathrm{constant}$, our calculations are performed 
fulfilling this relation. 
We will analyze the behavior of the gap for increasing $N$ 
by means of our exact diagonalization calculations up to $N =7$.

For the Laughlin state in the adiabatic/symmetric case, as is 
shown in Fig.~\ref{gap} for $N=4$ and $L=12$ there are 
two lowest excitations with $L=12-4=8$ and $L=13$ depending 
on the value of $\eta$. This is a general result for any 
$N$, the excitations have $L=N(N-1)-N$ and $L=N(N-1)+1$. 
The excitation with $L=13$ is an excitation of the center 
of mass of the system. This is due to the incompressibility 
of the Laughlin state. Namely, as shown in Fig.~\ref{yrast}, 
the state can increase its angular momentum 
without changing its interaction energy. We ignore this excitation, since 
we are interested in bulk excitations. This means that the 
linear left branch in the Laughlin region in Fig.~\ref{gap} must 
be extended up to $\eta=1$  (where one has the largest gap). All 
the branches on the right that would lie below this line, 
are edge excitations of different polarity.  In addition, the 
energy of the first excited state over the Laughlin denoted 
as $\Delta$, with $L=N(N-1)$ (i.e., in the same $L$-subspace) 
coincides with this largest gap at $\eta=1$ where effectively 
there is no trap.

\begin{figure*}[t]
\begin{center}
\includegraphics*[width=140mm]{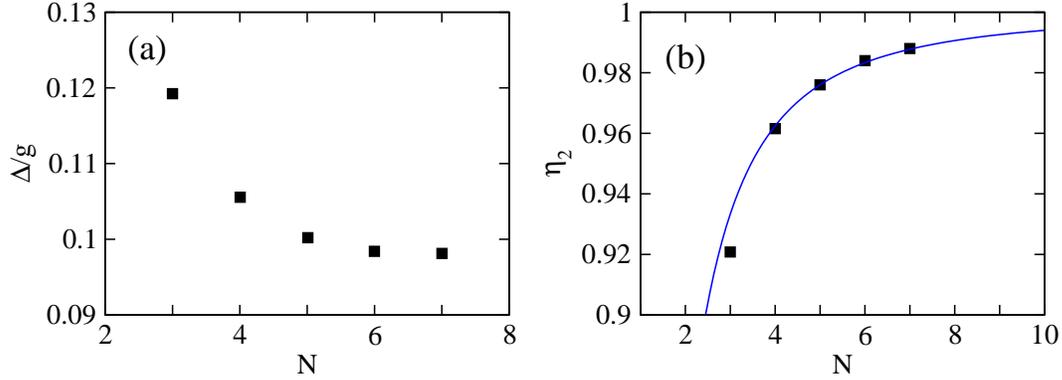}
\end{center}
\caption{(Color online)  
(a) $\Delta/g$, in units of $\hbar \omega_\perp$, of the Laughlin 
state as a function of $N$ in the adiabatic/symmetric case.
(b) Value of $\eta_2$ computed for $N=3, 4, 5, 6$, and $7$ in the 
adiabatic/symmetric case, compared to the prediction explained 
in the text, $\eta_2=1-0.1 (gN/N^2)$.}
\label{laugap}
\end{figure*}

As a consequence, to see the evolution of the gap over 
the Laughlin state it is sufficient to calculate the first 
two eigenvalues of the energy spectrum in the $L=N(N-1)$ 
subspace for each $N$. In Fig.~\ref{laugap}(a) we show the 
evolution of $\Delta/g$ with $N$. The tendency up to $N=7$ 
is to asymptotically recover the value of $0.1$ previously 
obtained by Regnault {\it et al.}~\cite{Regnault:2003} 
assuming a toroidal geometry and later reproduced by 
Roncaglia {\it et al.}~\cite{ron1}. Since we have taken 
$gN=6$, then $\Delta$ must compensate this $N$-dependence 
and consequently tend to zero as $\Delta \sim  1/N$. 
In effect, our results imply that the bosonic Laughlin state 
is observable only for few number of particles.

For possible practical implementations, it is also important 
to quantify the size of the parameter region where the Laughlin 
can be produced. Therefore, a good estimate of $\eta_2$ can 
be obtained taking advantage of the abovementioned coincidence. 
$\eta_2$ is the critical value where the energies of the Laughlin and 
$L=N(N-1)-N$ state cross, or 
\beq
(1-\eta_2) (L_0-N) \hbar\omega_\perp + V_1 =(1-\eta_2) L_0 \hbar\omega_\perp +V_0
\eeq
where $L_0=N(N-1)$ and $V_0$ and $V_1$ are $E_{\rm int}(L_0)$ and 
$E_{\rm int}(L_0-N)$ (see Fig.~\ref{yrast}), respectively. Or, 
\beq
\eta_2= 1 -{V_1\over N \hbar\omega_\perp} \qquad (V_0=0)\,.
\eeq
In addition, $V_1$ coincides with the energy difference 
$\Delta$ between the ground state and the first excitation 
in the $L_0$ subspace which tends, as $N$ increases 
(see Fig.~\ref{laugap})(a), to 
\beq
\Delta \sim 0.1 g \hbar \omega_\perp=0.1 (6/N) \hbar \omega_\perp
\eeq
and then 
\beq
\eta_2=1- \Delta / (N \hbar\omega_\perp)  \sim  1-0.1 {gN\over N^2}\,.
\eeq
In Fig.~\ref{laugap}(b) we compare the prediction of this formula 
and the computed $\eta_2$ for different $N$. As can be seen the formula 
agrees very well with the numerically obtained values.
 
\begin{figure*}[tbh]
\begin{center}
\includegraphics*[width=100mm]{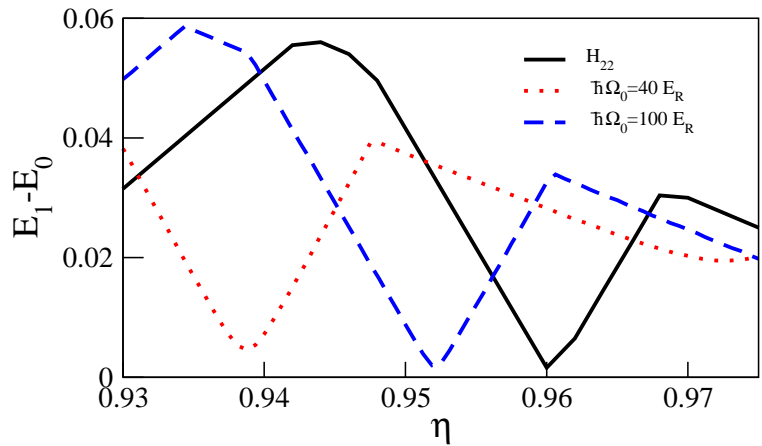}
\end{center}
\caption{(Color online) Evolution of the energy gap, in units 
of $\hbar \omega_\perp$ with decreasing $\Omega_0$ in the 
Laughlin region.}
\label{gapasym}
\end{figure*}

Finally, Fig.~\ref{gapasym} shows the change of the gap with decreasing 
$\Omega_0$. There are some important differences between the 
symmetric and the perturbed cases. The perturbation mixes a 
large number of subspaces and now it is not possible to ignore 
the right branch. The initial ($\eta_2$) and final frequencies 
at the boundaries of the Laughlin region are shifted to 
smaller values. The largest gap, the one at the upper vertex of 
the triangle is nearly constant. At $\eta_2$ the perturbation 
opens a gap where degeneracy occurs in the symmetric case. As a 
consequence, we conclude that on the one hand, the perturbation 
favors the observability, and on the other hand, the detection 
has to be restricted to a small number of particles.

\section{Summary and conclusions}
\label{sec7}

We have studied the possibility of producing relevant strongly 
correlated quantum states as ground states of a system of 
ultracold two-level atoms subjected to an artificial gauge 
field. The focus is on the formation of Moore-Read (Pfaffian), 
Laughlin and Laughlin-quasiparticle states. 
Considering a few number of atoms we have 
shown by exact diagonalization methods, that large squared 
overlaps between Pfaffian-like, Laughlin-quasiparticle-like 
and Laughlin-like states and the ground state of the system 
are found even for fairly small values of the external 
laser intensity $\hbar\Omega_0/E_R$. Reducing the laser 
intensity, correspondingly reducing the Rabi frequency, 
induces deformation in the system, increasing the angular 
momentum of the states. The structure of these deformed states 
is such that the main properties of the original undeformed 
states is retained in a broad region of parameter space. 
An analytical representation of the ground states consisting 
on the original states supplemented by a term affected by 
Jastrow factors increasing in two units their angular momentum provides 
large overlap with the numerical results, and allows to 
get analytical insights into the fractional behavior of 
the quasihole excitations. 

We have checked that quasihole states on the Laughlin and 
generalized Laughlin states can be produced by means of 
additional laser beams. We have studied the fractional 
charge and anyonic statistics of such quasiholes making 
use of the analytical representation of the states. Both, 
the effective charge and the statistical phase angle are 
close to the expected value of 1/2, even for the small 
number of particles considered here. Such fractional 
behavior is found whenever quasiholes are able to 
evolve in the bulk of the system, thus, the agreement with 
the expected behavior gets better as the size of the system 
is increased.  The admixture of higher angular momentum 
in the generalized Laughlin state does not modify the 
fractional behavior of its quasihole excitation. However, 
as the creation of two quasiholes requires a higher 
artificial field strength $\eta$ than the one necessary to get into the 
(generalized) Laughlin regime, our analysis is relevant mostly 
in the weakly perturbed regime, whereas in the highly perturbed 
regime an exact numerical evolution of the adiabatic movement 
would be needed, falling beyond the scope of the present work.  

Concerning the observability of Laughlin-like states, we find that 
decreasing $\hbar\Omega_0/E_R$ shifts the spectrum to smaller values 
of $\eta$, which are thus further away from the instability region 
$\eta=1$, favoring the experimental conditions. In addition, keeping 
the chemical potential constant, i.e. $N g=$ constant, we find that 
the observability of the Laughlin state is reduced to small number 
of particles due to the fast decrease of the largest possible gap 
over the ground state as $N$ increases.

An interesting aspect for future investigation are the excitations in 
the Pfaffian-like regime, which might obey non-Abelian statistics. 
A signature for this behavior would be a degeneracy in the state 
with four quasiholes~\cite{fradkin}. However, our present system is 
too small to test this physics.

\section*{Acknowledgments}
The authors thank N. Cooper, J. Dalibard and K. J. G\"unter for 
useful comments and discussions. This work has been supported by 
EU (NAMEQUAM, AQUTE, MIDAS), ERC (QUAGATUA), 
Spanish MINCIN (FIS2008-00784, FIS2010-16185, 
FIS2008-01661 and QOIT Consolider-Ingenio 2010), 
Alexander von Humboldt Stiftung, IFRAF and ANR (BOFL project). 
B.~J.-D. is supported by a Grup Consolidat SGR 21-2009-2013. 
M. L. acknowledges Hamburg Theory Award.
\clearpage
\appendix

\section{Effective Hamiltonian}
\label{apb}

Let us consider a system described by a single particle Hamiltonian 
given by $H=H_0+V$ where $H_0$ is solvable and $V$ 
can be treated as a perturbation. In addition, let us assume that 
the eigenvalues of $H_0$ are grouped in manifolds well-separated 
in anergy, i.e., the eigenenergies $E_{i\alpha}$ have the following 
property,
\begin{equation}
|E_{i\alpha}-E_{j\alpha}| \ll |E_{i\alpha}-E_{j\beta}|
\end{equation}
where $\alpha$ and $\beta$ denote different manifolds and the 
index $i$ denotes different states inside a manifold. In our 
case we have two manifolds, the lowest energy one ($\alpha=2$) 
described by $H_{22}$ and the most energetic one described by 
$H_{11}$ ($\beta=1$), both together play the role of $H_0$,
\beq
 H_{0}=
\left(\matrix{
H_{11} & 0\cr
0& H_{22}
}\right)
\eeq
acting on the spinor shown below Eq.(~\ref{eq6}). 

The difference in energy, $\hbar \Omega_0$, is much larger than 
the typical expected values of $H_{22}$ which are of the order of the 
recoil energy $E_R=\hbar^2 k^2/2M)$. Within the lowest manifold 
we make the LLL assumption. This scenario is valid in the case where 
the two internal states are uncoupled and the system evolves always 
in $|\psi_1\rangle$ or $|\psi_2\rangle$. 

Instead, if there is 
coupling between the manifolds, the Hamiltonian is represented 
by $H_0+\lambda V$ where $\lambda$ is a dimensionless parameter 
and $V$, in general, has non-zero matrix elements inside each 
manifold as well as between them. As long as $\lambda$ is small, 
the structure of the well-separated manifolds and their degeneracy 
is preserved with slight modifications. Physically, the coupling 
between the two manifolds means that the motion of an atom in 
$|\psi_2\rangle$ is slightly modified by a sudden and short 
time period in the other manifold. The high frequency dynamics 
driven by $(E_{i\alpha}-E_{i \beta})/\hbar$ ($\alpha\ne \beta$) 
is averaged by the slow dynamics driven by $(E_{i\alpha}-E_{j\alpha})/\hbar$. 
In a way, the wave functions of the manifold-2 are ``influenced'' 
(or dressed) by the wave functions of manifold-1. In our case, 
the structure of $V$ is, 
\beq
V=
\left(\matrix{
0 & H_{12}\cr
H_{21}& 0
}\right)\,,
\eeq
with no diagonal elements.

The main goal will be to obtain an effective Hermitian Hamiltonian 
$H'$ that acts only on the unperturbed manifold-2, though having 
the same eigenvalues as those of $H$, and with zero matrix elements 
between the two unperturbed manifolds. In this way, we will be 
able to consider only the FD functios and ignore manifold-1.

The Hermiticity and the coincidence on the eigenvalues with $H$ 
are achieved if we consider a unitary transformation from $H$ 
to $H'$ as,
\begin{equation}
H'=T \,H \,T^{\dag}
\end{equation}
where $T=e^{iS}$ and $S=S^{\dag}$.
Or
\begin{eqnarray}
H'&=&e^{iS} H e^{-iS} \nonumber \\
&=&H+[iS,H]+\frac{1}{2!}
\left[iS,\left[iS,H\right]\right]+ \dots= H_0+\lambda H'_1
+\lambda^2 H'_2+ \dots
\end{eqnarray}
where we have considered the expansion $S=\lambda S_1+\lambda^2 S_2+ \dots$ 
and the condition that $S$ has zero matrix elements inside each manifold. 
Grouping the terms in orders of $\lambda$, it is possible to solve $S_i$ 
step by step as functions of known quantities. One arrives to the expression:
\bea
\sand{i\alpha}{H'}{j\alpha} &=&E_{i\alpha}\delta_{ij}
+\sand{i\alpha}{\lambda V}{j\alpha}
+\frac{1}{2}\sum_{k,\gamma\ne \alpha} \sand{i\alpha}{\lambda
  V}{k\gamma}\sand{k\gamma}{\lambda V}{j\alpha} \nonumber \\
&\times &\left[\frac{1}{E_{i\alpha}-E_{k\gamma}}+\frac{1}{E_{j\alpha}-E_{k\gamma}}\right] + \dots
\eea
The first term represents the unperturbed levels in the manifold-2, 
the second one is the direct coupling between unperturbed levels in 
the manifold-2 and the third term is the contribution of the indirect 
coupling through the manifold-1. In our case the last equation reduces 
to,
\beq
\sand{i\alpha}{H'}{j\alpha}\simeq
E_{i2}\delta_{ij}+
\frac{\lambda^2}{2}\sum_{k}\sand{i2}{V}{k1}\sand{k1}{V}{j2}
\left(-\frac{1}{\Omega_0} -\frac{1}{\Omega_0}\right)
\eeq
where we have approximated $E_{k1}$ by $\hbar\Omega_0$ and considered $E_{i2}\ll\hbar\Omega_0$
or
\beq
\sand{i\alpha}{H'}{j\alpha}=E_{i2}\delta_{ij}-\frac{\lambda^2}{\Omega_0}\sand{i2}{V^2}{j2}
\eeq
thus,
\beq
H'=H_{22}-\frac{H_{21}H_{12}}{\Omega_0}
\eeq
which is the results used in Eq.~(\ref{eq10}) in Section~\ref{sec2}. The
interaction term is considered part of the non-perturbed term together 
with the kinetic contribution, see Eq.~(\ref{eq9}).

The explicit form of the perturbation term $H_{21}H_{12}$ used is, 
\bea
H_{21}H_{12} &=& 
\left(\frac{\hbar ^4}{4 M^2 w^4}-\frac{2 x^2 \hbar ^4}{M^2 w^6}+\frac{k^2 x^2
  \hbar ^4}{16 M^2 w^4}+\frac{k^4 x^2 \hbar ^4}{64 M^2 w^2}\right.\nonumber \\
&&\left.+\frac{i k x y
  \hbar ^4}{4 M^2 w^5}+\frac{k^2 y^2 \hbar ^4}{64 M^2 w^4}\right) \nonumber \\
&&+\left(-\frac{i k x \hbar ^4}{4 M^2 w^3}
-\frac{i k^3 x \hbar ^4}{8 M^2  w}\right) \partial_y 
+\left(\frac{x \hbar ^4}{M^2 w^4}
-\frac{i k y \hbar ^4}{8 M^2 w^3}\right) \partial_x \nonumber \\
&&+\left(-\frac{k^2 \hbar ^4}{4 M^2}+\frac{k^2 x^2 \hbar ^4}{4 M^2 w^2}\right)
\partial^2_y+\left(-\frac{\hbar ^4}{4 M^2 w^2}+\frac{x^2 \hbar ^4}{2 M^2 w^4}\right)
\partial^2_x \,.
\eea
one can show that this operator does not conserve $L$, as it connects 
$L'$-subspaces with $L'=L+\Delta$ where $\Delta=0,\pm 2$, $\pm 4$.

\end{document}